\documentclass{aastex62}
\usepackage{longtable}
\usepackage{booktabs}
\usepackage{lineno}
\newcommand{\psr}{PSR~J2032+4127}
\received{}
\revised{}
\accepted{}
\submitjournal{ApJ}

\shorttitle{X-Rays from PSR~J2032+4127 in 2017}
\shortauthors{Pal et al.}
\begin{document}

\title{X-Ray spectral evolution of PSR~J2032+4127 during the 2017 periastron passage}

\correspondingauthor{P.~H.~T.~Tam}
\email{tanbxuan@mail.sysu.edu.cn}

\author[0000-0001-8922-8391]{Partha Sarathi Pal}
\affiliation{School of Physics \& Astronomy, Sun Yat Sen University, \\ 135 West Xingang Road, Guangzhou, 510275, China}
\author[0000-0002-1262-7375]{P.~H.~T.~Tam}
\affiliation{School of Physics \& Astronomy, Sun Yat Sen University, \\ 135 West Xingang Road, Guangzhou, 510275, China}
\author{Y.~Cui}
\affiliation{School of Physics \& Astronomy, Sun Yat Sen University, \\ 135 West Xingang Road, Guangzhou, 510275, China}
\author[0000-0002-0439-7047]{K.~L.~Li}
\affiliation{Department of Physics, Ulsan National Institute of Science and Technology, \\ Ulsan, 44919, Korea}
\author{A.~K.~H.~Kong}
\affiliation{Astrophysics, Department of Physics, University of Oxford, \\ Keble Road, Oxford OX1 3RH, UK}
\affiliation{Institute of Astronomy \& Department of Physics, National Tsing Hua University, \\ Hsinchu 30013, Taiwan}
\author[0000-0003-3791-3754]{C.~G\"{u}ng\"{o}r}
\affiliation{Faculty of Engineering and Natural Science, Sabanc{\i} University, \\ Orhanli - Tuzla, Istanbul, 34956, Turkey}

%\linenumbers
\begin{abstract}
We report X-ray data analysis results obtained from {\it Chandra}, {\it XMM-Newton}, {\it NuSTAR} and {\it Swift}
observations of \psr~taken before, during, and after the periastron on 2017 November 13.
We found the first clear evidence of a change in the X-ray spectral index over
the passage period, thanks to a broad and sensitive spectral coverage by {\it XMM-Newton} and {\it NuSTAR}.
We analysed the joint {\it XMM-Newton} and {\it NuSTAR} observation epochs with power-law and 
broken power-law model. We have obtained change in spectral parameters before and after the periastron passage 
for both models. The spectra get softened after the passage. The evolution of the spectral index and break energy 
before and after the periastron may indicate a change in the physical state of shock-accelerated electrons.
\end{abstract}

\section{Introduction} \label{sec:intro}
\psr~is a radio-loud GeV emitting pulsar, discovered by {\it Fermi Large Area Telescope (Fermi-LAT)}
in a blind search of {\it Fermi-LAT} gamma-ray data \citep{abdo09} and was subsequently detected 
in radio wavelength with the Green Bank Telescope \citep[GBT;][]{camilo09}.
It is one of the younger member of its class and has a spin period of 143~ms \citep{abdo13}.  
Several observations revealed a variable spin-down rate in the pulsar which 
is explained as an effect of Doppler shift due to the pulsar's motion along the 
orbit of the binary system with long-period \citep{lyne15}. The companion star was identified 
as a Be star, MT~91-213, which has a mass of around 15 $M_\sun$ with a circumstellar
variable disk within radius 0.2-0.5 AU \citep{ho17}. The pulsar spin-down luminosity(\.{E}) was 
determined to be around $1.7 \times 10^{35}~erg~s^{-1}$ with a characteristic age of 180 kyr, and 
the binary system is at a distance of 1.44 kpc in the Cyg OB2 stellar cluster \citep{gaia18}.
Further observations precisely determined the orbital parameters with a binary 
period of 45-50 years, an eccentricity between 0.94 and 0.99 and a longitude of
periastron between 21\arcdeg and 52\arcdeg.   
According to the best fit of radio timing solution the periastron passage was on 2017 November 13 
with an approximate separation between \psr~and MT~91-213 of 1~AU \citep{ho17, coe17}.

Close to the periastron, due to the expected interaction between the matter of the stellar wind 
(or possibly the disk) and the pulsar wind, emission of low energy (radio to X-rays) to high energy 
(GeV or up to TeV) photons are predicted \citep{takata17, bednarek18}. 
Indeed, this object shows similarity with another $\gamma$-ray binary PSR~B1259-63/LS~2883 
\citep{chernya15, tam15, j18, tam18}. PSR~B1259-63 has a spin period of 47.8 ms and orbits around the
 Be star LS~2883 with a period of 3.4 years in a highly eccentric orbit (e $\sim$ 0.87).
It shows a spin down luminosity around $8.3 \times 10^{35}~erg~s^{-1}$.
It is known that PSR B1259-63 converts nearly all of its pulsar spin-down luminosity to GeV 
emission during parts of its orbit \citep{cali15}. 
Several models have been proposed to explain the non-thermal, broadband radiation from
PSR~B1259-63 as well as the GeV flares.
These include synchrotron emission and/or inverse-Compton (IC) up-scattering off stellar
photons or disk photons from the companion star by the accelerated electrons in the shock
between the PW and stellar wind \citep{tavani97, kirk99, dubus06, bogo08, khan11, takata12}, 
unshocked PW particles \citep{khan12}, and Doppler boosting effects \citep{dubus10, takata12}.
As it is accepted that both systems are similar \citep{ho17, coe19}, it is understandable that 
the origin of X-ray \& TeV emission for \psr~should be pulsar spin-down luminosity.

\citet{ho17} and \citet{li17a} first reported the X-ray brightening from the direction of \psr, 
consistent with the increased shock emission. \citet{ho17} has predicted the timing of the 
periastron from orbital solution.
\citet{li17c} has summarized {\it The Neil Gehrels Swift Observatory (Swift)}, 
{\it the Chandra X-ray Observatory (Chandra)}, 
{\it The Nuclear Spectroscopic Telescope Array Mission (NuSTAR)} and 
{\it The European Space Agency's (ESA) X-ray Multi-Mirror Mission (XMM-Newton)} 
observations of \psr~up to early 2017 and tried to explain the pre-periastron X-ray behaviors of \psr. 
In continuation of this work, \citet{li18} reported the X-ray modulation using Swift data and $\gamma$-ray 
observations of \psr~over the 2017 periastron. The effect of an asymmetric stellar wind with polar gradient is 
investigated in \citet{petro18}.
Optical variability around the periastron is monitored and reported by \citet{kolka17}.  
The VHE gamma-ray counterpart of the 2017 periastron emission was observed by
{\it Very Energetic Radiation Imaging Telescope Array System (VERITAS)} and 
{\it Major Atmospheric Gamma Imaging Cherenkov (MAGIC)} \citep{ab18}.
TeV spectral analysis data is adopted in this paper for comparison with X-Ray analysis result.
\citet{takata17} explained the multi-wavelength phenomena in light of the shock scenario and discussed 
the possible formation of an accretion disk. However, the rather rich broad-band data and the complex X-ray 
flux variability still lacks a full theoretical understanding.

In this paper, we present the result throughout our analysis of the archival data of {\it Swift, XMM-Newton, Chandra} 
and {\it NuSTAR} to understand the detailed X-ray evolution through the periastron passage.
In particular, we report the first accurate X-ray spectral index and interstellar absorption 
measurements over the 2017 periastron passage of \psr.

\section{Data Analysis} \label{sec:data}

During 2017 periastron \psr~was observed by {\it Swift}, {\it XMM-Newton}(P.I: Jules Halpern), 
{\it Chandra}(P.I.: Jules Halpern), {\it NuSTAR}(P.I: Jules Halpern). 
The observation details is shown in Table.~\ref{table1}. 

\begin{longtable}{cccccc}
\caption{Observational data analysed in this paper. First column shows the observation dates. Next column shows the observation epoch w.r.t periastron date (MJD 58070). In third column shows the observing instrument. Fourth column shows observation ID. Fifth column shows the exposure time. Sixth column shows count rate during the observation. \label{table1}}\\\toprule
\hline
Date & Epoch & Instrument & Obs-Id& Exposure & Rate \\
(MJD) & (Days) & & & Sec & counts/s \\
\hline
\endfirsthead
\multicolumn{6}{@{}l}{\ldots continued}\\\hline
\hline
Date & Epoch & Instrument & Obs-Id& Exposure & Rate \\
(MJD) & (Days) & & & Sec & counts/s \\
\hline
\endhead
\hline
\multicolumn{6}{r@{}}{continued \ldots}\\
\endfoot
\bottomrule
\endlastfoot
57442.34 & -628.16 & {\it Chandra} & 18788 & 4896 & 0.03$\pm$0.003 \\ 
57785.20 & -285.30 & {\it Chandra} & 19607 & 26640 & 0.07$\pm$0.002 \\ 
57901.66 & -168.84 & {\it Chandra} & 19700 & 28160 & 0.15$\pm$0.002 \\ 
57976.33 & -94.17 & {\it SWIFT} & 93146011 & 3336 & 0.04$\pm$0.003 \\ 
57987.51 & -82.99 & {\it SWIFT} & 34282090 & 1166 & 0.04$\pm$0.006 \\ 
57990.10 & -80.40 & {\it SWIFT} & 34282091 & 1992 & 0.03$\pm$0.004 \\ 
57991.36 & -79.14 & {\it SWIFT} & 93146012 & 2651 & 0.04$\pm$0.004 \\ 
57994.08 & -76.42 & {\it Chandra} & 19701 & 28400 & 0.13$\pm$0.002 \\ 
58002.05 & -68.45 & {\it SWIFT} & 93148006 & 2804 & 0.04$\pm$0.004 \\ 
58007.22 & -63.28 & {\it SWIFT} & 93148007 & 1146 & 0.03$\pm$0.006 \\ 
58017.00 & -53.50 & {\it SWIFT} & 93148008 & 2437 & 0.05$\pm$0.004 \\ 
58018.38 & -52.12 & {\it SWIFT} & 34282094 & 1613 & 0.06$\pm$0.006 \\ 
58019.91 & -50.59 & {\it SWIFT} & 93146014 & 1455 & 0.04$\pm$0.007 \\ 
58026.15 & -44.35 & {\it SWIFT} & 93146015 & 2956 & 0.04$\pm$0.004 \\ 
58032.59 & -37.91 & {\it SWIFT} & 93148009 & 2614 & 0.05$\pm$0.004 \\ 
58033.32 & -37.18 & {\it SWIFT} & 93146016 & 3221 & 0.04$\pm$0.004 \\ 
58039.11 & -31.39 & {\it SWIFT} & 34282097 & 1418 & 0.05$\pm$0.006 \\ 
58039.32 & -31.18 & {\it SWIFT} & 93148010 & 3576 & 0.05$\pm$0.004 \\ 
58046.09 & -24.41 & {\it SWIFT} & 34282098 & 1988 & 0.05$\pm$0.005 \\ 
58047.08 & -23.42 & {\it SWIFT} & 93146018 & 2710 & 0.05$\pm$0.005 \\ 
58047.35 & -23.15 & {\it SWIFT} & 93148011 & 2083 & 0.03$\pm$0.004 \\ 
58049.48 & -21.02 & {\it SWIFT} & 88016001 & 1756 & 0.04$\pm$0.005 \\ 
58049.51 & -20.99 & {\it NuSTAR} & 30302002002 & 37843 & 0.06$\pm$0.001 \\ 
58049.74 & -20.76 & {\it XMM} & 801910201 & 22643 & 0.43$\pm$0.005 \\ 
58050.74 & -19.76 & {\it SWIFT} & 34282099 & 2894 & 0.04$\pm$0.004 \\ 
58051.13 & -19.37 & {\it SWIFT} & 93148012 & 1893 & 0.04$\pm$0.005 \\ 
58053.00 & -17.50 & {\it SWIFT} & 34282100 & 1096 & 0.04$\pm$0.006 \\ 
58054.92 & -15.58 & {\it SWIFT} & 93146019 & 2789 & 0.04$\pm$0.004 \\ 
58057.51 & -12.99 & {\it SWIFT} & 34282101 & 1626 & 0.03$\pm$0.005 \\ 
58060.25 & -10.25 & {\it Chandra} & 19702 & 19080 & 0.12$\pm$0.003 \\ 
58060.64 & -9.86 & {\it SWIFT} & 34282103 & 1981 & 0.03$\pm$0.004 \\ 
58061.04 & -9.46 & {\it SWIFT} & 93146020 & 3328 & 0.03$\pm$0.003 \\ 
58062.20 & -8.30 & {\it NuSTAR} & 90302321002 & 39826 & 0.03$\pm$0.001 \\ 
58062.76 & -7.74 & {\it SWIFT} & 93148013 & 3219 & 0.02$\pm$0.003 \\ 
58062.78 & -7.72 & {\it XMM} & 801910301 & 16657 & 0.22$\pm$0.004 \\ 
58062.89 & -7.61 & {\it Chandra} & 20836 & 9560 & 0.10$\pm$0.003 \\ 
58065.41 & -5.09 & {\it SWIFT} & 34282106 & 1848 & 0.02$\pm$0.004 \\ 
58068.14 & -2.36 & {\it SWIFT} & 93146021 & 2349 & 0.03$\pm$0.004 \\ 
58069.78 & -0.72 & {\it NuSTAR} & 30302002004 & 42705 & 0.03$\pm$0.001 \\ 
58069.87 & -0.63 & {\it SWIFT} & 34282108 & 1526 & 0.02$\pm$0.004 \\ 
58070.50 & 0.00 & {\it XMM} & 801910401 & 11518 & 0.21$\pm$0.005 \\ 
58071.85 & 1.35 & {\it SWIFT} & 34282109 & 2333 & 0.01$\pm$0.003 \\ 
58074.65 & 4.15 & {\it SWIFT} & 34282110 & 1953 & 0.03$\pm$0.004 \\ 
58075.58 & 5.08 & {\it SWIFT} & 93146022 & 2817 & 0.02$\pm$0.003 \\ 
58076.10 & 5.60 & {\it SWIFT} & 34282111 & 2250 & 0.02$\pm$0.003 \\ 
58077.50 & 7.00 & {\it SWIFT} & 93148015 & 3366 & 0.04$\pm$0.003 \\ 
58078.51 & 8.01 & {\it NuSTAR} & 90302321004 & 40003 & 0.06$\pm$0.001 \\ 
58078.96 & 8.46 & {\it SWIFT} & 34282112 & 1483 & 0.05$\pm$0.006 \\ 
58079.05 & 8.55 & {\it XMM} & 801910501 & 17958 & 0.49$\pm$0.006 \\ 
58081.35 & 10.85 & {\it SWIFT} & 34282114 & 1491 & 0.04$\pm$0.005 \\ 
58082.21 & 11.71 & {\it SWIFT} & 93146023 & 1274 & 0.04$\pm$0.006 \\ 
58083.14 & 12.64 & {\it SWIFT} & 34282115 & 1973 & 0.04$\pm$0.005 \\ 
58084.14 & 13.64 & {\it SWIFT} & 93148016 & 2172 & 0.05$\pm$0.005 \\ 
58085.00 & 14.50 & {\it SWIFT} & 34282116 & 2317 & 0.05$\pm$0.005 \\ 
58087.21 & 16.71 & {\it SWIFT} & 34282117 & 2028 & 0.04$\pm$0.005 \\ 
58089.13 & 18.63 & {\it SWIFT} & 93146024 & 2130 & 0.05$\pm$0.005 \\ 
58091.46 & 20.96 & {\it NuSTAR} & 30302002006 & 41531 & 0.04$\pm$0.001 \\ 
58091.72 & 21.22 & {\it XMM} & 801910601 & 16756 & 0.36$\pm$0.006 \\ 
58092.73 & 22.23 & {\it SWIFT} & 93148017 & 3122 & 0.04$\pm$0.004 \\ 
58095.38 & 24.88 & {\it SWIFT} & 34282119 & 1465 & 0.02$\pm$0.004 \\ 
58095.44 & 24.94 & {\it SWIFT} & 10451001 & 3459 & 0.00$\pm$0.002 \\ 
58127.99 & 57.49 & {\it Chandra} & 19608 & 29030 & 0.11$\pm$0.002 \\ 
58193.59 & 123.09 & {\it Chandra} & 19698 & 29160 & 0.04$\pm$0.001 \\ 
\end{longtable}

{\it XMM-Newton} have observed \psr~on 5 different epochs between MJD 58049 to 58091 
with an average 17 ks exposure. 
{\it XMM-Newton EPIC-PN} and {\it MOS} data are used for the X-ray analysis in this paper.
The data reduction were reprocessed with the software 
\texttt{SASv16.1\footnote{\url{https://www.cosmos.esa.int/web/xmm-newton/sas-threads}}}, 
using the most updated calibration files (updated to 2018 May) \citep{str01, tur01}. 
The {\it EPIC-PN} event files were reprocessed from observation 
data files with \texttt{epproc} with `bad'(e.g., `hot',`dead', `flickering') pixels removed. 
The period of high background events were examined and excluded using the light curves in 
the energy band 10-12 keV. As the X-ray of \psr~is not bright and there is no pile-up effect, 
we extracted the source events from a circular region of radius 20\arcsec, 
using single and double events (PATTERN$\leq$4, FLAG=0). The background events were made from a source-free 
circular region of radius of 20\arcsec. We grouped the pn spectra to have at least 25 counts in each bin, 
so as to adopt the $\chi^2$ statistic for the spectral fitting. 
The ancillary response files (arfs) were extracted with {\it arfgen}. The {\it EPIC-PN}
redistribution matrix file was extracted with {\it rmfgen} and used in spectral fitting.
Similarly {\it EPIC-MOS} event files were reprocessed from observation 
data files with \texttt{emproc} with `bad'(e.g., `hot',`dead', `flickering') pixels removed.
The period of high background events were examined and excluded using the light curves in 
the energy band above 10 keV. We checked for pile-up and the pile-up effect 
is not serious in the source center. We extracted the source events from a circular region of radius 20\arcsec, 
using single and double events (PATTERN$\leq$12). We grouped the pn spectra to have at least 25 counts in each bin, 
so as to adopt the $\chi^2$ statistic for the spectral fitting. The ancillary response files (arfs) were extracted 
with {\it arfgen}. We repeated the similar extraction method for {\it MOS1} and {\it MOS2} spectra.
During fitting we used both of them as different data groups. 

{\it NuSTAR} have observed \psr~almost simultaneously with {\it XMM-Newton} enabling us to perform a joint 
spectral analysis for {\it XMM} and {\it NuSTAR}. The {\it NuSTAR} event files are obtained with \texttt{nupipeline} 
of \texttt{HEASOFTv6.25\footnote{\url{https://heasarc.gsfc.nasa.gov/docs/software/heasoft/}}} software \citep{har13}.
We have used {\it SW\_SAA=0} and {\it SW\_TENTACLE=0} during analysis to avoid background flaring.
With \texttt{nuproduct} the source spectra are extracted from a circular region of 30\arcsec as the source position and 
the background is estimated from a source free region of same size. The spectra are grouped with 30 counts per bin for 
spectral analysis. The response functions and ancillary respond files are also extracted. Both {\it FPMA} and {\it FPMB}
spectra are used for spectral analysis. During fitting we used both FPMA and FPMB as different data groups. 
    
{\it Chandra} have observed \psr~on 5 different epochs around 2017 periastron between MJD 57785 and MJD 58062
with average 20 ks exposure. The data reduction is performed with \texttt{CIAOv4.9\footnote{\url{http://cxc.harvard.edu/ciao/}}} 
software \citep{wei00}. The fits files are cleaned for high flaring backgrounds. The spectra are extracted from 
a circular region of 7\arcsec at the source position and background is estimated from a source free region with 
same radius from the cleaned fits file. We grouped {\it ACIS} spectra to have at least 15 counts in each bin for spectral 
analysis. The PI had chosen a smaller good exposure time in the proposal to avoid the pile-up effect. We have 
tested the spectra with \texttt{XSPEC} pile-up model but there is no pile-up effect observed in the spectra. 

For {\it Swift-XRT} data reduction, the level 2 cleaned event files of {\it Swift-XRT} 
are obtained from the events of photon counting (PC) mode data with \texttt{xrtpipeline} \citep{burr05}. 
The spectra are extracted from a circular region in the best source position with 20\arcsec radius. 
The background is estimated from an annular region in the same position with radii from 30\arcsec to 60\arcsec.
The {\it ancillary response files (arfs)} are extracted with \texttt{xrtmkarf}. The PC 
{\it redistribution matrix file (rmf) version (v.12)} is used in the spectral fits.

\begin{figure}[ht!]
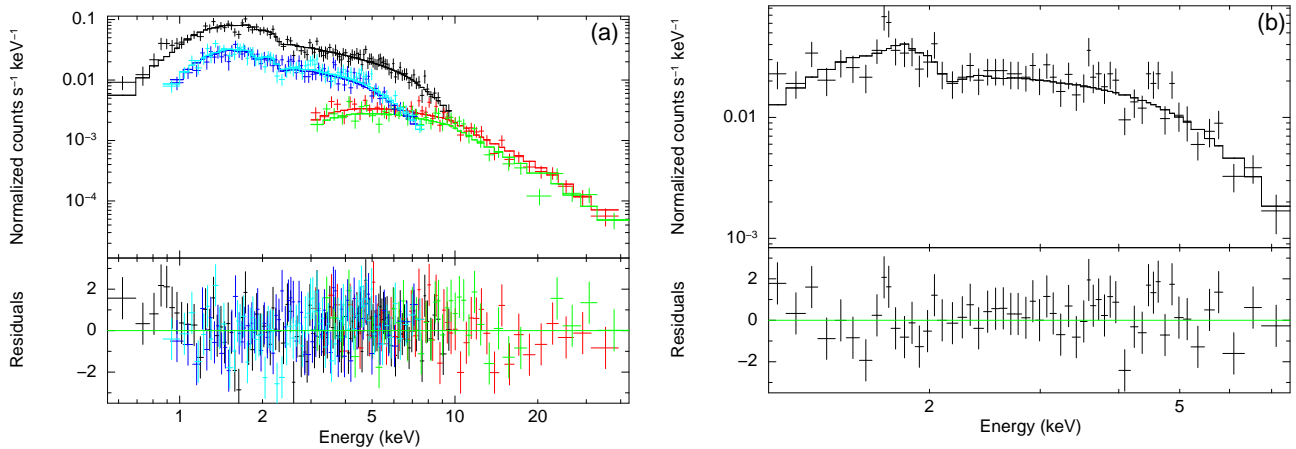

\includegraphics[scale=0.325,angle=270]{58062_fit.eps}
\includegraphics[scale=0.325,angle=270]{20836_fit.eps}
\caption{(a) Joint {\it XMM-Newton} and {\it NuSTAR} absorbed power-law fitted spectra of MJD 58062. 
The black curve represents {\it XMM-Newton EPIC-PN} data. Blue and cyan curves represent {\it XMM-Newton EPIC-MOS1} 
and {\it MOS2} data respectively. Red and green curves represent the {\it NuSTAR FPMA} and {\it FPMB} data respectively. 
(b) {\it Chandra-ACIS} absorbed power-law fitted spectrum of MJD 58062. 
\label{spec}}
\end{figure}

The spectral analysis is performed with \texttt{XSPEC(v12.10.1)\footnote{\url{https://heasarc.gsfc.nasa.gov/xanadu/xspec/}}}.
We applied the absorption model ({\it tbabs} in {\it XSPEC}) to account for interstellar effect due to hydrogen 
column density \citep{wam00}. 
Although the Galactic hydrogen column density of the position of \psr~is given as 
$(1.2-1.5)\times10^{22}\mathrm{cm^{-2}}$ by \citet{dl90, kal05}, we kept the parameter free while 
fitting to investigate whether there exists any significant evolution in the 
absorption parameter during the periastron. We fit the spectrum using the power law model; 
tbabs $\times$ po. The uncertainties are given at 90\% confidence levels for one parameter.
The joint spectral analysis of {\it XMM} and {\it NuSTAR} spectral analysis is done 
in the 0.5-50~keV energy range. We multiplied the model with a constant parameter to better 
calibrate different detectors. The constant kept fixed to unity for the {\it XMM EPIC-PN} 
spectrum and remained free for {\it EPIC-MOS1, MOS2, FPMA, FPMB}.
On MJD 58062 the 0.5-50~keV joint {\it XMM+NuSTAR} spectral analysis show the interstellar absorption 
around $0.77\pm0.06\times10^{22}\mathrm{cm^{-2}}$ along with the power-law index $\Gamma=1.37\pm0.05$
and 0.3-10~keV unabsorbed power-law flux is around $0.23\pm0.004\times10^{-11}~erg~s^{-1}\mathrm{cm^{-2}}$. 
The cross-calibration constant for {\it FPMA} is obtained as $1.02\pm0.07$ and for {\it FPMB} is $0.88\pm0.07$
In Fig.~\ref{spec}(a), {\it XMM+NuSTAR} fitted spectra of MJD 58062 are shown.
The {\it XMM+NuSTAR} spectral analysis results are shown in detail in Table.~\ref{table2}.
The cross-calibration constants for different instruments, during the joint {\it XMM+NuSTAR} spectral analysis with absorbed 
power-law model, are also reported in Table.~\ref{table2}. Most of the cases the cross-calibration constant values are
around unity. For a few cases larger deviations are obtained. May be they are due to the presence of stray light
or some other effects \citep{wik14}.  

The {\it Chandra ACIS} spectra analyzed with absorbed power-law model within 0.8-10 keV.
from the spectral analysis the interstellar absorption is obtained around 
$1.06\pm0.38\times10^{22}\mathrm{cm^{-2}}$ along with the power-law index $\Gamma=1.29\pm0.28$
and 0.3-10~keV unabsorbed power-law flux is around $0.29\pm0.017\times10^{-11}~erg~s^{-1}\mathrm{cm^{-2}}$. 
The fitted {\it ACIS} spectrum of MJD 58062 is shown in Fig.~\ref{spec}(b). 
The {\it Chandra} spectral analysis results are shown in detail in Table.~\ref{table3}.

\begin{deluxetable}{cccccccccc}
\tablecaption{Spectral analysis results for joint observations of {\it XMM-Newton} and {\it NuSTAR} with absorbed power-law model. 
First column shows the observation dates. In second column interstellar absorption values are shown. Third column shows the 
power-law index. Fourth column shows the 0.3-10 keV unabsorbed flux calculated from analysis results. Fifth column shows 
the reduced $\chi^2$ values with degrees of freedom. Next sixth to tenth columns show the cross-calibration constants for 
different instruments. The cross-calibration constant for {\it EPIC-PN} is kept fixed to unity during the analysis. \label{table2}}
\tablehead{
\colhead{Date} & \colhead{$n_H$} & \colhead{$\Gamma$} & \colhead{Flux} & \colhead{$\chi_{\nu}^2$} 
& \multicolumn{3}{c}{\it XMM-Newton} & \multicolumn{2}{c}{\it NuSTAR}\\
\colhead{(MJD)} & \colhead{($10^{22}cm^{-2}$)} & \colhead{} & \colhead{($10^{-12}erg~s^{-1}\mathrm{cm^{-2}}$)} & \colhead{(dof)} 
& \colhead{\it PN}& \colhead{\it MOS1} & \colhead{\it MOS2} & \colhead{\it FPMA} & \colhead{\it FPMB}}
\startdata
58049 & $0.85\pm0.04$ & $1.38\pm0.03$ & $4.66\pm0.05$ & 1.09(463)& 1.0 & $0.97\pm0.03$ & $1.00\pm0.03$ & $0.96\pm0.05$ & $0.94\pm0.05$ \\ 
58062 & $0.77\pm0.06$ & $1.37\pm0.05$ & $2.31\pm0.04$ & 1.13(319)& 1.0 & $1.02\pm0.05$ & $1.08\pm0.05$ & $1.02\pm0.07$ & $0.88\pm0.07$ \\ 
58070 & $0.68\pm0.06$ & $1.37\pm0.06$ & $2.00\pm0.04$ & 1.10(238)& 1.0 & $0.98\pm0.06$ & $1.05\pm0.07$ & $0.91\pm0.07$ & $0.85\pm0.07$ \\ 
58079 & $0.78\pm0.03$ & $1.53\pm0.03$ & $5.01\pm0.06$ & 1.24(445)& 1.0 & $1.00\pm0.03$ & $1.02\pm0.03$ & $1.00\pm0.05$ & $0.94\pm0.05$ \\ 
58091 & $0.81\pm0.05$ & $1.63\pm0.05$ & $3.52\pm0.06$ & 1.04(350)& 1.0 & $0.98\pm0.05$ & $1.04\pm0.05$ & $0.98\pm0.06$ & $1.04\pm0.07$ \\ 
\enddata
\end{deluxetable}

\begin{deluxetable}{ccccc}
\tablecaption{Spectral analysis results for {\it Chandra} observations with absorbed power-law model. First column shows the observation dates. In second column interstellar absorption values are shown. Third column shows the power-law index. Fourth column shows the 0.3-10 keV unabsorbed flux calculated from analysis results. Fifth column shows the reduced $\chi^2$ values with degrees of freedom. \label{table3}}
\tablehead{
\colhead{Date} & \colhead{$n_H$} & \colhead{$\Gamma$} & \colhead{Flux} & \colhead{$\chi_{\nu}^2$} \\
\colhead{(MJD)} & \colhead{($10^{22}cm^{-2}$)} & \colhead{} & \colhead{($10^{-12}erg~s^{-1}\mathrm{cm^{-2}}$)} & \colhead{(dof)}
}
\startdata
57442 & $ 0.54\pm1.27$ & $ 1.93\pm1.36$ & $ 0.66\pm0.12$ & 1.75(4) \\ 
57785 & $ 1.02\pm0.18$ & $ 2.03\pm0.18$ & $ 2.04\pm0.08$ & 0.97(91) \\ 
57901 & $ 0.97\pm0.12$ & $ 1.73\pm0.12$ & $ 4.09\pm0.11$ & 0.74(189) \\ 
57994 & $ 0.86\pm0.13$ & $ 1.45\pm0.12$ & $ 3.50\pm0.10$ & 0.89(179) \\ 
58060 & $ 0.87\pm0.20$ & $ 1.19\pm0.16$ & $ 3.28\pm0.12$ & 1.12(119) \\ 
58062 & $ 1.06\pm0.38$ & $ 1.29\pm0.28$ & $ 2.90\pm0.17$ & 1.11(54) \\ 
58127 & $ 1.07\pm0.12$ & $ 1.68\pm0.12$ & $ 3.24\pm0.10$ & 0.90(156) \\ 
58193 & $ 0.85\pm0.22$ & $ 1.76\pm0.24$ & $ 0.95\pm0.05$ & 0.92(59) \\ 
\enddata
\end{deluxetable}
%57442 & {\it Chandra} & $ 0.54\pm1.27$ & $ 1.93\pm1.36$ & $ 0.66\pm0.12$ & 1.75(4) \\ 
%57785 & {\it Chandra} & $ 1.02\pm0.18$ & $ 2.03\pm0.18$ & $ 2.04\pm0.08$ & 0.97(91) \\ 
%57901 & {\it Chandra} & $ 0.97\pm0.12$ & $ 1.73\pm0.12$ & $ 4.09\pm0.11$ & 0.74(189) \\ 
%57994 & {\it Chandra} & $ 0.86\pm0.13$ & $ 1.45\pm0.12$ & $ 3.50\pm0.10$ & 0.89(179) \\ 
%58049 & {\it XMM+NuSTAR} & $ 0.85\pm0.04$ & $ 1.38\pm0.03$ & $ 4.66\pm0.05$ & 1.09(463) \\ 
%58060 & {\it Chandra} & $ 0.87\pm0.20$ & $ 1.19\pm0.16$ & $ 3.28\pm0.12$ & 1.12(119) \\ 
%58062 & {\it Chandra} & $ 1.06\pm0.38$ & $ 1.29\pm0.28$ & $ 2.90\pm0.17$ & 1.11(54) \\ 
%58062 & {\it XMM+NuSTAR} & $ 0.77\pm0.06$ & $ 1.37\pm0.05$ & $ 2.31\pm0.04$ & 1.13(319) \\ 
%58070 & {\it XMM+NuSTAR} & $ 0.68\pm0.06$ & $ 1.37\pm0.06$ & $ 2.00\pm0.04$ & 1.10(238) \\ 
%58079 & {\it XMM+NuSTAR} & $ 0.78\pm0.03$ & $ 1.53\pm0.03$ & $ 5.01\pm0.06$ & 1.24(445) \\ 
%58091 & {\it XMM+NuSTAR} & $ 0.81\pm0.05$ & $ 1.63\pm0.05$ & $ 3.52\pm0.06$ & 1.04(350) \\ 
%58127 & {\it Chandra} & $ 1.07\pm0.12$ & $ 1.68\pm0.12$ & $ 3.24\pm0.10$ & 0.90(156) \\ 
%58193 & {\it Chandra} & $ 0.85\pm0.22$ & $ 1.76\pm0.24$ & $ 0.95\pm0.05$ & 0.92(59) \\ 

From {\it XMM+NuSTAR} spectral analysis, the average $n_H$ value is obtained around $n_H=0.78\times10^{22}\mathrm{cm^{-2}}$.
Here all 1.0-10.0 keV {\it Swift-XRT} spectra are fitted with the same model component with
 fixed $n_H=0.78\times10^{22}\mathrm{cm^{-2}}$. XRT unabsorbed flux values are shown in Fig.~\ref{7panel} for comparison.
In all cases the unabsorbed power-law flux is calculated in 0.3-10 keV energy range in cgs units.

We have calculated the hardness ratio for the X-Ray data. Here hardness ratio is the
ratio between 2.0-10.0 keV photons and 0.3-2.0 keV photons.

For {\it Swift-UVOT} data reduction, all extensions of sky images are stacked with \texttt{uvotimsum}.
The source magnitudes are derived with 3-$\sigma$ significance level      
from the circular region of 5\arcsec radius in the best source position
of the stacked sky images from all the filters with \texttt{uvotsource}. The background is estimated from an 
annular region in the same position with radii from 10\arcsec to 20\arcsec.

During {\it XMM+NuSTAR} joint observations the correlation between interstellar absorption and 
power-law index is calculated with 50 steps in both parameters within the error-bar of the parameter. 
Then 68\%, 90\% and 99\% confidence contours are plotted with red, blue and gray curves in Fig.~\ref{cont}.    

During the {\it XMM+NuSTAR} joint observations we have also performed spectral fit with absorbed broken power-law
model. %Spectral parameters are shows in Table.~\ref{table4}. 
The fitted spectra for MJD 58070 with both absorbed power-law and broken power-law model are plotted in Fig.~\ref{bkn} 
for comparison.

\section{Results} \label{sec:result}

\begin{figure}[ht!]
\includegraphics[scale=0.6,angle=270]{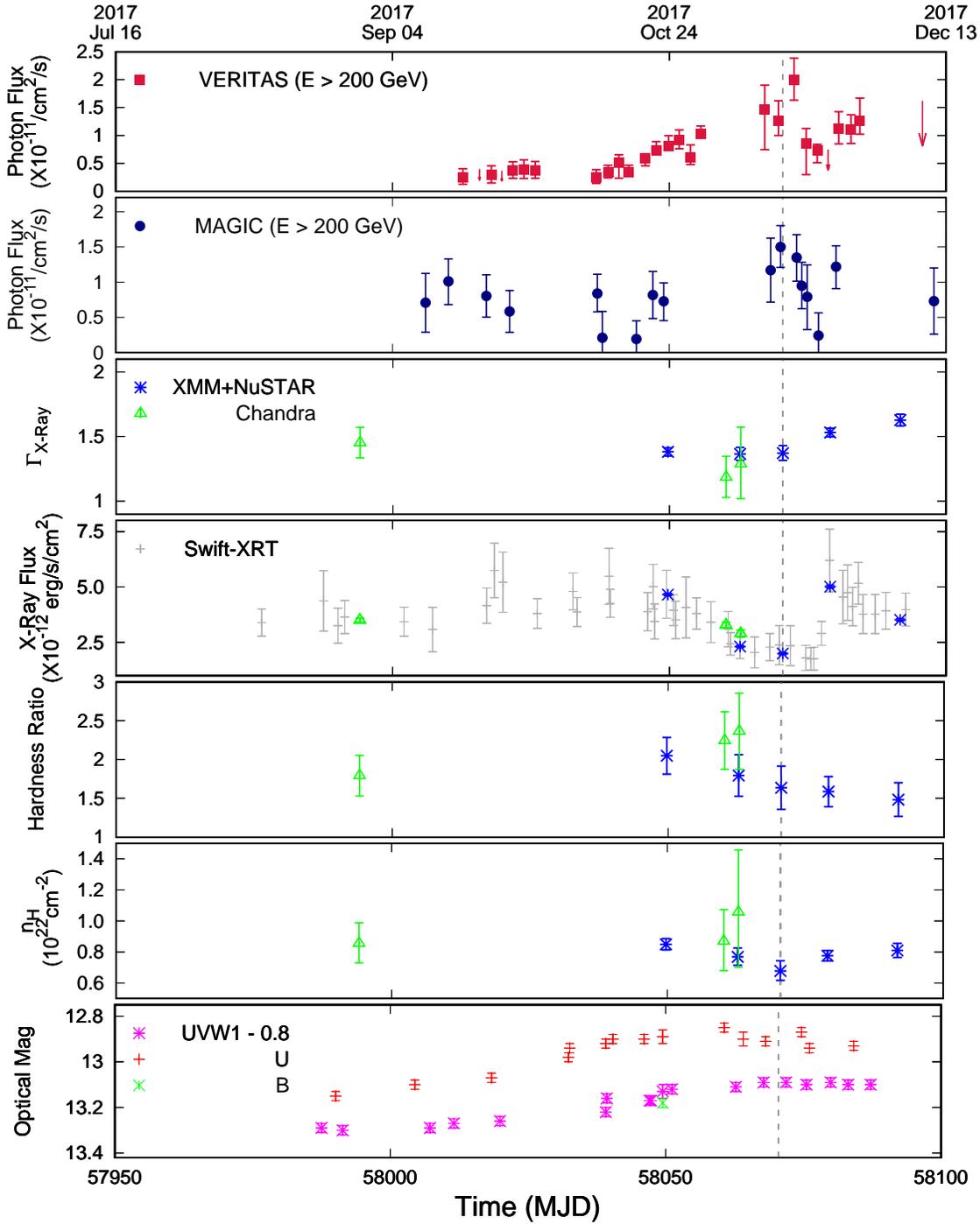}
\caption{Top panel shows the {\it VERITAS} photon flux light curve. Second 
panel shows the {\it MAGIC} photon flux light curve. {\it VERITAS} and {\it MAGIC} (E $>$ 200 GeV) 
data points are obtained from lower panel of Fig.~1(b) of \citet{ab18}. 
Third panel shows the variation of X-ray power law index. In the fourth, fifth and sixth panel
the Blue stars represent {\it XMM+NuSTAR} result. Green triangle represents {\it Chandra} result. 
Fourth panel shows the 0.3-10 keV unabsorbed power-law flux variation where gray points represent 
{\it Swift-XRT} data analysis result. Fifth panel shows the evolution of hardness ratio between 
2.0-10.0 keV and 0.3-2.0 keV photons. Sixth panel shows the interstellar absorption obtained from spectral fitting. 
The bottom panel shows the {\it UVOT} magnitude in Vega system. Red points represent the U magnitude.
Green point represent the B magnitude and pink star represent UVW1 magnitude with offset by 0.8
magnitude. The grey dashed line for all panels represents the periastron on MJD 58070.
From third to sixth panel during the X-Ray analysis uncertainties are given at 90\% confidence levels. 
\label{7panel}}
\end{figure}

We analyze all pre-periastron X-ray data and confirm results published previously. Here we report the results 
of the spectral analysis of new {\it Chandra}, {\it XMM-Newton}, {\it NuSTAR} and {\it SWIFT} data.
In 7 panel plot of the Fig.~\ref{7panel} we have plotted the analysis result and compared with the 
TeV analysis results obtained from \citet{ab18}. 

\begin{figure}[ht!]
\includegraphics[scale=0.55,angle=270]{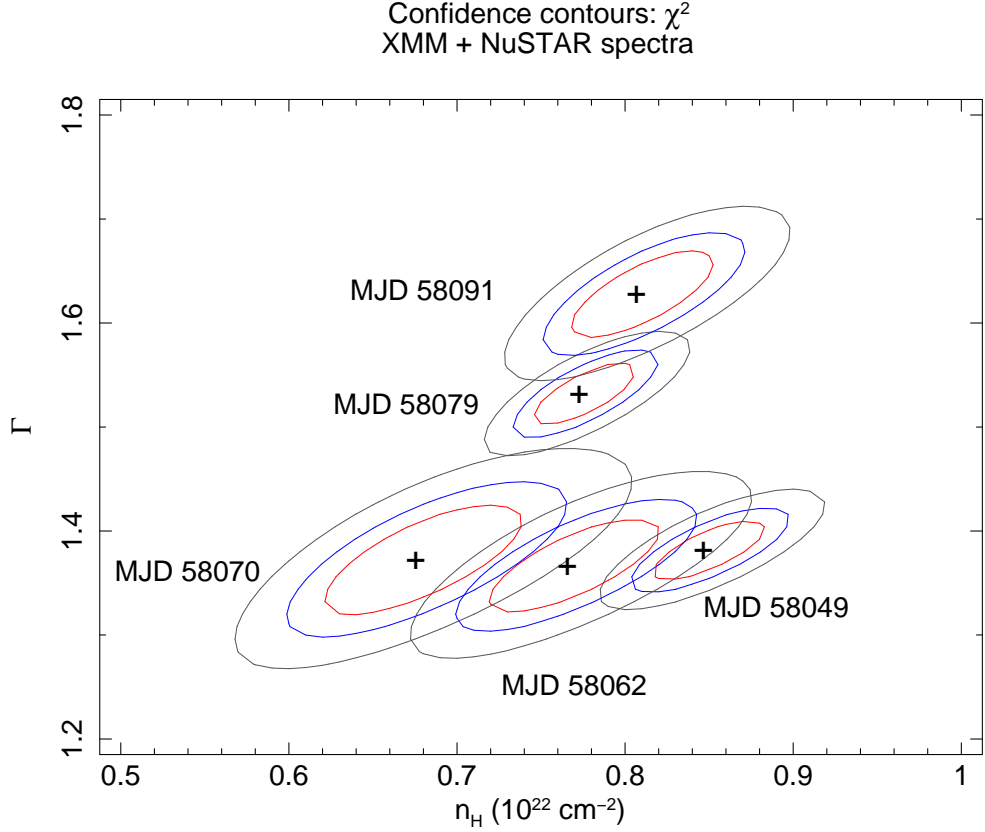}
\caption{
Evolution of $\chi^2$ confidence contours between interstellar absorption 
and power-law index obtained from {\it XMM} and {\it NuSTAR} spectral analysis with power-law model.
Here 68\% 90\% and 99\% confidence contours are represented with red, blue and
gray curves respectively.
\label{cont}}
\end{figure}

In the top and second panel of Fig.~\ref{7panel} the TeV photon flux values are reported 
from {\it VERITAS} and {\it MAGIC} observations respectively. These data points are obtained from 
lower panel Fig.~1(b) of \citet{ab18} for comparison. 
In the top panel the from MJD 58012 to MJD 58042 the {\it VERITAS} (E $>$ 200 GeV) photon
counts are below $0.5\times10^{-11}~Counts~\mathrm{cm^{-2}s^{-1}}$ after MJD 58045 the {\it VERITAS} counts
started increasing gradually to $1.5\times10^{-11}~Counts~\mathrm{cm^{-2}s^{-1}}$ just before the periastron.
After the periastron it sharply increase to $2\times10^{-11}~Counts~\mathrm{cm^{-2}s^{-1}}$ and sharply 
decrease to $0.75\times10^{-11}~Counts~\mathrm{cm^{-2}s^{-1}}$ and then the count rate remain variable 
around $1.0\times10^{-11}~Counts~\mathrm{cm^{-2}s^{-1}}$.
In the second panel the {\it MAGIC} (E $>$ 200 GeV) count rates remain varying around 
$0.75\times10^{-11}~Counts~\mathrm{cm^{-2}s^{-1}}$ during the pre-periastron time span. 
On periastron the count rate is on its peak at $1.5\times10^{-11}~Counts~\mathrm{cm^{-2}s^{-1}}$.
Then it decreases to $0.25\times10^{-11}~Counts~\mathrm{cm^{-2}s^{-1}}$ within 7 days after the periastron. 
In the third panel the X-ray power-law index is reported. The {\it XMM+NuSTAR} power-law index 
remain steady around $1.38 \pm 0.03$ from MJD 58049 to MJD 58070, the periastron. Then the spectrum gets softer gradually
from $1.37 \pm 0.03$ to $1.63 \pm 0.05$ up to MJD 58091. This is shown with blue stars in the plot. 
The {\it Chandra} analysis results are also shown in the plot with green triangles for comparison. 
In the fourth panel the 0.3-10 keV unabsorbed X-ray power-law flux evolution with time is shown. 
The {\it XMM+NuSTAR} flux is shown with blue stars. {\it Chandra} flux is reported with Green triangles. 
The grey points represent the {\it Swift-XRT} flux values. From MJD 58050 to 58069 the flux
decreases from $(4.66 \pm 0.05)\times 10^{-12}~erg~s^{-1}\mathrm{cm^{-2}}$ 
to $(2.0 \pm 0.04)\times 10^{-12}~erg~s^{-1}\mathrm{cm^{-2}}$ 
upto its periastron then it increases to $(5.01 \pm 0.06)\times 10^{-12}~erg~s^{-1}\mathrm{cm^{-2}}$ then decreases to 
$(3.52 \pm 0.06)\times 10^{-12}~erg~s^{-1}\mathrm{cm^{-2}}$, confirming the double-hump structure found previously. 
We have calculated the hardness ratio between 2.0-10.0 keV photons and 0.3-2.0 keV photons. 
The fifth panel show the hardness ratio evolution of the X-ray data. The {\it XMM-PN} data is shown with blue stars which shows a 
gradual decrease in hardness ratio from $2.04 \pm 0.24$ to $1.48 \pm 0.22$. {\it Chandra} data is reported with Green triangles. 
In the sixth panel we show the variation of interstellar absorption variation obtained from fitting.
Here we only put {\it Chandra}(Green triangle) and {\it XMM+NuSTAR}(Blue star) data. 
Albeit with large error bars we note that within measurement uncertainties $n_H$ is consistent with no significant 
change within 99\% confidence level (see Fig.~\ref{cont}) and the best-fit $n_H$ values are slightly smaller than 
the Galactic $n_H$ values \citep{dl90, kal05}, but similar to the foreground value estimated by the optical color 
excess of MT~91-213 (i.e. $n_H = 0.77\times10^{22}\mathrm{cm^{-2}}$) \citep{camilo09,li17c}.
%In Table.~\ref{table2} the spectral analysis parameters for different observations are shown in details.
In the bottom panel we have shown the {\it UVOT} variation during the periastron. Here we can see 
a U band magnitude remained constant around 13 magnitude (Vega) and UVW1 filter magnitude 
remain constant around 14 Mag(Vega). This optical variation is believed to be observed due to
optical evolution from the companion star.    

In Fig.~\ref{cont} we show the evolution of the 68\%(red), 90\%(blue) and 99\%(gray) 
confidence contours between $n_H$ and $\Gamma$ obtained from the {\it XMM+NuSTAR} analysis (assuming power-law).

Any high energy exponential cutoff may represent a signature of an accretion disk around the compact object.
We also analyzed the {\it XMM+NuSTAR} observation with absorbed power-law with exponential
cutoff model and we did not observe any presence of cut-off in the observed data. 
This result confirms that there are no evidence of an accretion disc during periastron, 
similar to other gamma-ray binaries \citep{hasi09, an15, tam15, li14}.

\begin{figure}[ht!]
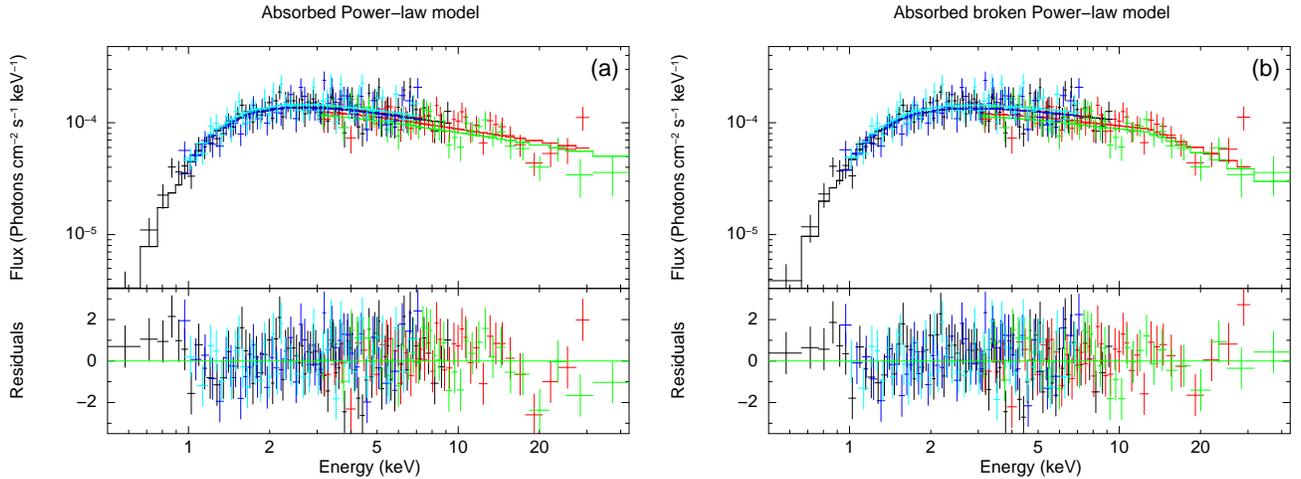

\includegraphics[scale=0.325,angle=270]{58070_po.eps}
\includegraphics[scale=0.325,angle=270]{58070_bkn.eps}
\caption{(a) Joint {\it XMM-Newton} and {\it NuSTAR} absorbed power-law fitted spectra of MJD 58070. 
(b) Joint {\it XMM-Newton} and {\it NuSTAR} absorbed broken power-law fitted spectra of MJD 58070. 
The black curve represents {\it XMM-Newton EPIC-PN} data. Blue and cyan curves represent {\it XMM-Newton EPIC-MOS1} 
and {\it MOS2} data respectively. Red and green curves represent the {\it NuSTAR FPMA} and {\it FPMB} data respectively. 
\label{bkn}}
\end{figure}

\citet{u09} has reported, presence of a power-law break around $5.0 \pm 0.7~keV$ for PSR~B1259-63 over the periastron passage,
as magnetic interaction between shock and pulsar wind.
We have tried to fit {\it XMM+NuSTAR} spectra with absorbed broken power-law model. 
%and the result is reported in Table.~\ref{table4}. 
With broken power-law model we have obtained similar goodness of fit as power-law model for all five epochs. 
Here we also found a break in power-law, by $\Delta\Gamma \sim 0.35$ around $\varepsilon_{br} \sim 5~keV$, 
during pre-periastron period and increases to $\varepsilon_{br} = 12.9 \pm 3.9~keV$ by $\Delta\Gamma \simeq 0.7$
exactly at the periastron passage (see Fig.~\ref{bkn}). May be this is due to some magnetic properties of wind. 
The relation between break energy and magnetization parameter of wind for PSR~B1259-63 is given by: 
\begin{equation}
\varepsilon_{br} \simeq 4\left(\frac{B}{1.8~G}\right)\left(\frac{\gamma_1}{4\times10^5}\right)^2 keV
\end{equation}
\citep[Eq.~9;][]{u09}.
If we assume \psr~and PSR~B1259-63 are similar sources and Lorentz factor($\gamma_1 \sim 4\times10^5$) for 
pulsar wind then during periastron passage, $12.9 \pm 3.9~keV$ break energy will represent pulsar wind magnetic field 
around $5.8 \pm 1.7~G$ for \psr.
Interestingly, we found no $n_H$ change at all in the broken power-law model. 
Finally, we note that no short-term evolution is observed within each {\it XMM-Newton}, 
{\it NuSTAR} and {\it Chandra} observation, indicating that the spectral changes happen in longer time scales.

\section{Discussion \& Conclusion} \label{sec:discussion}

The X-ray flux derived from the more accurate X-ray measurements presented in this work is in excellent agreement 
with those obtained based on {\it Swift-XRT} only, which has more coverage in time albeit with large errors  
\citep{ho17, li17c, li18, petro18}.
A model light curve developed by~\citet{takata17} and~\citet{li18} generally matches the {\it Swift-XRT} light curve reasonably well, 
including the pre-periastron X-ray peak at MJD~58020, and the abrupt drop just before the periastron. 
As discussed in~\citet{li18}, it is not obvious how the flare-like X-ray emission observed around MJD 58080-58100 is 
created in the wind-wind interaction model. 
Furthermore, with the detailed X-ray spectral analysis, we could deliver the evolution of the intrinsic X-ray PL index.
Instead, the pulsar wind and Be stellar disk interaction may enhance 
the X-ray flux, if (1) the scale height of the Be stellar disk at the pulsar position is larger than the shock radius, 
such that the disk can confine most of the pulsar wind, enhancing the pulsar wind dissipation, and (2) a relatively 
large base density of the Be stellar disk~\citep{li18}. One of our major results, namely, a softer intrinsic spectral
 index (around 1.6) at the post-periastron flare time as compared to that during and before periastron (around 1.4), 
is in line with a possible new emission region/mechanism at the flare time.
The emission mechanism of X-ray photons are driven by Synchrotron emission for binary systems without stellar 
disc i.e.~LS~5039 and 1FGL~J1018.6-5856 \citep{hasi09,an15}. 
We note that for PSR~B1259-63/LS~2883 system, a softer X-ray power-law index is clearly seen and is thought to 
be related to the disc passage. Synchrotron radiation is responsible for the emission of X-ray photons in this 
system \citep{chen19}. \citet{coe19} has reported the presence of circumstellar disk and the interaction of the 
compact object with the circumstellar disk for \psr~during the periastron. 
Now we can infer that the Inverse Compton Scattering may be the cause of the spectral softening for \psr~after 
periastron passage.

The Power-law index has shown a clear softening after the periastron, which may indicate the variation in physical state of the 
shock-accelerated electrons. The fit with broken power-law model also gives similar goodness of fit as we have obtained from 
power-law model. The increase in break energy from around 5 keV before periastron to around 13 keV at periastron may 
indicate some physical change during the periastron passage. 
We do not know the exact cause of the possible changing column density, it can be due to the stellar 
disk or other components of the stellar winds.

\acknowledgments
PSP, YC acknowledges sysu-postdoctoral fellowship. 
PSP, PHT, and YC are supported by NSFC through grants 
11633007, 11661161010, and U1731136. We acknowledge the
referee for advice to improve the manuscript. 

\vspace{5mm}
\facilities{XMM-Newton, Swift(XRT and UVOT), CXO, NuSTAR}
\software{HEASOFT \citep[v6.25;][]{heasarc14}, 
          XSPEC \citep[v12.10.1;][]{xspec96}, 
          CIAO \citep[v4.9;][]{ciao06}, 
          SAS \citep[v16.1;][]{sas04}}
%\software{astropy \citep{2013A&A...558A..33A},  
%          Cloudy \citep{2013RMxAA..49..137F}, 
%          SExtractor \citep{1996A&AS..117..393B}
%          }

\bibliographystyle{aasjournal}
\bibliography{ref} % if your bibtex file is called example.bib

\begin{thebibliography}{}
\expandafter\ifx\csname natexlab\endcsname\relax\def\natexlab#1{#1}\fi
\providecommand{\url}[1]{\href{#1}{#1}}
\providecommand{\dodoi}[1]{doi:~\href{http://doi.org/#1}{\nolinkurl{#1}}}
\providecommand{\doeprint}[1]{\href{http://ascl.net/#1}{\nolinkurl{http://ascl.net/#1}}}
\providecommand{\doarXiv}[1]{\href{https://arxiv.org/abs/#1}{\nolinkurl{https://arxiv.org/abs/#1}}}

\bibitem[{{Abdo} {et~al.}(2009){Abdo}, {Ackermann}, {Ajello}, {Atwood},
  {Axelsson}, {Baldini}, {Ballet}, {Barbiellini}, {Baring}, {Bastieri},
  {Baughman}, {Bechtol}, {Bellazzini}, {Berenji}, {Bignami}, {Blandford},
  {Bloom}, {Bonamente}, {Borgland}, {Bregeon}, {Brez}, {Brigida}, {Bruel},
  {Burnett}, {Caliandro}, {Cameron}, {Camilo}, {Caraveo}, {Carlson},
  {Casandjian}, {Cecchi}, {{\c C}elik}, {Charles}, {Chekhtman}, {Cheung},
  {Chiang}, {Ciprini}, {Claus}, {Cognard}, {Cohen-Tanugi}, {Cominsky},
  {Conrad}, {Corbet}, {Cutini}, {Dermer}, {Desvignes}, {de Angelis}, {de Luca},
  {de Palma}, {Digel}, {Dormody}, {do Couto e Silva}, {Drell}, {Dubois},
  {Dumora}, {Edmonds}, {Farnier}, {Favuzzi}, {Fegan}, {Focke}, {Frailis},
  {Freire}, {Fukazawa}, {Funk}, {Fusco}, {Gargano}, {Gasparrini}, {Gehrels},
  {Germani}, {Giebels}, {Giglietto}, {Giordano}, {Glanzman}, {Godfrey},
  {Grenier}, {Grondin}, {Grove}, {Guillemot}, {Guiriec}, {Hanabata}, {Harding},
  {Hayashida}, {Hays}, {Hobbs}, {Hughes}, {J{\'o}hannesson}, {Johnson},
  {Johnson}, {Johnson}, {Johnson}, {Johnston}, {Kamae}, {Katagiri}, {Kataoka},
  {Kawai}, {Kerr}, {Kn{\"o}dlseder}, {Kocian}, {Kramer}, {Kuss}, {Lande},
  {Latronico}, {Lemoine-Goumard}, {Longo}, {Loparco}, {Lott}, {Lovellette},
  {Lubrano}, {Madejski}, {Makeev}, {Manchester}, {Marelli}, {Mazziotta},
  {McConville}, {McEnery}, {McLaughlin}, {Meurer}, {Michelson}, {Mitthumsiri},
  {Mizuno}, {Moiseev}, {Monte}, {Monzani}, {Morselli}, {Moskalenko}, {Murgia},
  {Nolan}, {Norris}, {Nuss}, {Ohsugi}, {Omodei}, {Orlando}, {Ormes}, {Paneque},
  {Panetta}, {Parent}, {Pelassa}, {Pepe}, {Pesce-Rollins}, {Piron}, {Porter},
  {Rain{\`o}}, {Rando}, {Ransom}, {Ray}, {Razzano}, {Rea}, {Reimer}, {Reimer},
  {Reposeur}, {Ritz}, {Rochester}, {Rodriguez}, {Romani}, {Roth}, {Ryde},
  {Sadrozinski}, {Sanchez}, {Sander}, {Saz Parkinson}, {Scargle}, {Schalk},
  {Sgr{\`o}}, {Siskind}, {Smith}, {Smith}, {Spandre}, {Spinelli}, {Stappers},
  {Starck}, {Striani}, {Strickman}, {Suson}, {Tajima}, {Takahashi}, {Tanaka},
  {Thayer}, {Thayer}, {Theureau}, {Thompson}, {Thorsett}, {Tibaldo}, {Torres},
  {Tosti}, {Tramacere}, {Uchiyama}, {Usher}, {Van Etten}, {Vasileiou},
  {Venter}, {Vilchez}, {Vitale}, {Waite}, {Wallace}, {Wang}, {Watters}, {Webb},
  {Weltevrede}, {Winer}, {Wood}, {Ylinen}, \& {Ziegler}}]{abdo09}
{Abdo}, A.~A., {Ackermann}, M., {Ajello}, M., {et~al.} 2009, Science, 325, 848,
  \dodoi{10.1126/science.1176113}

\bibitem[{{Abdo} {et~al.}(2013){Abdo}, {Ajello}, {Allafort}, {Baldini},
  {Ballet}, {Barbiellini}, {Baring}, {Bastieri}, {Belfiore}, {Bellazzini}, \&
  et~al.}]{abdo13}
{Abdo}, A.~A., {Ajello}, M., {Allafort}, A., {et~al.} 2013, \apjs, 208, 17,
  \dodoi{10.1088/0067-0049/208/2/17}

\bibitem[{{Abeysekara} {et~al.}(2018){Abeysekara}, {Benbow}, {Bird}, {Brill},
  {Brose}, {Buckley}, {Chromey}, {Daniel}, {Falcone}, {Finley}, \&
  et~al.}]{ab18}
{Abeysekara}, A.~U., {Benbow}, W., {Bird}, R., {et~al.} 2018, \apjl, 867, L19,
  \dodoi{10.3847/2041-8213/aae70e}

\bibitem[{{An} {et~al.}(2015){An}, {Bellm}, {Bhalerao}, {Boggs}, {Christensen},
  {Craig}, {Fuerst}, {Hailey}, {Harrison}, {Kaspi}, {Natalucci}, {Stern},
  {Tomsick}, \& {Zhang}}]{an15}
{An}, H., {Bellm}, E., {Bhalerao}, V., {et~al.} 2015, \apj, 806, 166,
  \dodoi{10.1088/0004-637X/806/2/166}

\bibitem[{{Arnaud}(1996)}]{xspec96}
{Arnaud}, K.~A. 1996, in Astronomical Society of the Pacific Conference Series,
  Vol. 101, Astronomical Data Analysis Software and Systems V, ed. G.~H.
  {Jacoby} \& J.~{Barnes}, 17

\bibitem[{{Bednarek} {et~al.}(2018){Bednarek}, {Banasi{\'n}ski}, \&
  {Sitarek}}]{bednarek18}
{Bednarek}, W., {Banasi{\'n}ski}, P., \& {Sitarek}, J. 2018, Journal of Physics
  G Nuclear Physics, 45, 015201, \dodoi{10.1088/1361-6471/aa97ee}

\bibitem[{{Bogovalov} {et~al.}(2008){Bogovalov}, {Khangulyan}, {Koldoba},
  {Ustyugova}, \& {Aharonian}}]{bogo08}
{Bogovalov}, S.~V., {Khangulyan}, D.~V., {Koldoba}, A.~V., {Ustyugova}, G.~V.,
  \& {Aharonian}, F.~A. 2008, \mnras, 387, 63,
  \dodoi{10.1111/j.1365-2966.2008.13226.x}

\bibitem[{{Burrows} {et~al.}(2005){Burrows}, {Hill}, {Nousek}, {Kennea},
  {Wells}, {Osborne}, {Abbey}, {Beardmore}, {Mukerjee}, {Short}, {Chincarini},
  {Campana}, {Citterio}, {Moretti}, {Pagani}, {Tagliaferri}, {Giommi},
  {Capalbi}, {Tamburelli}, {Angelini}, {Cusumano}, {Br{\"a}uninger}, {Burkert},
  \& {Hartner}}]{burr05}
{Burrows}, D.~N., {Hill}, J.~E., {Nousek}, J.~A., {et~al.} 2005, \ssr, 120,
  165, \dodoi{10.1007/s11214-005-5097-2}

\bibitem[{{Caliandro} {et~al.}(2015){Caliandro}, {Cheung}, {Li}, {Scargle},
  {Torres}, {Wood}, \& {Chernyakova}}]{cali15}
{Caliandro}, G.~A., {Cheung}, C.~C., {Li}, J., {et~al.} 2015, \apj, 811, 68,
  \dodoi{10.1088/0004-637X/811/1/68}

\bibitem[{{Camilo} {et~al.}(2009){Camilo}, {Ray}, {Ransom}, {Burgay},
  {Johnson}, {Kerr}, {Gotthelf}, {Halpern}, {Reynolds}, {Romani}, {Demorest},
  {Johnston}, {van Straten}, {Saz Parkinson}, {Ziegler}, {Dormody}, {Thompson},
  {Smith}, {Harding}, {Abdo}, {Crawford}, {Freire}, {Keith}, {Kramer},
  {Roberts}, {Weltevrede}, \& {Wood}}]{camilo09}
{Camilo}, F., {Ray}, P.~S., {Ransom}, S.~M., {et~al.} 2009, \apj, 705, 1,
  \dodoi{10.1088/0004-637X/705/1/1}

\bibitem[{{Chen} {et~al.}(2019){Chen}, {Takata}, {Yi}, {Yu}, \&
  {Cheng}}]{chen19}
{Chen}, A.~M., {Takata}, J., {Yi}, S.~X., {Yu}, Y.~W., \& {Cheng}, K.~S. 2019,
  arXiv e-prints, arXiv:1904.07527.
\newblock \doarXiv{1904.07527}

\bibitem[{{Chernyakova} {et~al.}(2015){Chernyakova}, {Neronov}, {van Soelen},
  {Callanan}, {O'Shaughnessy}, {Babyk}, {Tsygankov}, {Vovk}, {Krivonos},
  {Tomsick}, {Malyshev}, {Li}, {Wood}, {Torres}, {Zhang}, {Kretschmar},
  {McSwain}, {Buckley}, \& {Koen}}]{chernya15}
{Chernyakova}, M., {Neronov}, A., {van Soelen}, B., {et~al.} 2015, \mnras, 454,
  1358, \dodoi{10.1093/mnras/stv1988}

\bibitem[{{Coe} {et~al.}(2017){Coe}, {Steele}, {Ho}, {Stappers}, {Lyne},
  {Halpern}, {Ray}, {Johnson}, {Ng}, \& {Kerr}}]{coe17}
{Coe}, M.~J., {Steele}, I.~A., {Ho}, W.~C.~G., {et~al.} 2017, The Astronomer's
  Telegram, 10920

\bibitem[{{Coe} {et~al.}(2019){Coe}, {Okazaki}, {Steele}, {Ng}, {Ho}, {Lyne},
  {Stappers}, {Johnson}, {Ray}, \& {Kerr}}]{coe19}
{Coe}, M.~J., {Okazaki}, A.~T., {Steele}, I.~A., {et~al.} 2019, \mnras, 485,
  1864, \dodoi{10.1093/mnras/stz515}

\bibitem[{{Dickey} \& {Lockman}(1990)}]{dl90}
{Dickey}, J.~M., \& {Lockman}, F.~J. 1990, \araa, 28, 215,
  \dodoi{10.1146/annurev.aa.28.090190.001243}

\bibitem[{{Dubus}(2006)}]{dubus06}
{Dubus}, G. 2006, \aap, 451, 9, \dodoi{10.1051/0004-6361:20054233}

\bibitem[{{Dubus} {et~al.}(2010){Dubus}, {Cerutti}, \& {Henri}}]{dubus10}
{Dubus}, G., {Cerutti}, B., \& {Henri}, G. 2010, \aap, 516, A18,
  \dodoi{10.1051/0004-6361/201014023}

\bibitem[{{Fruscione} {et~al.}(2006){Fruscione}, {McDowell}, {Allen},
  {Brickhouse}, {Burke}, {Davis}, {Durham}, {Elvis}, {Galle}, \&
  {Harris}}]{ciao06}
{Fruscione}, A., {McDowell}, J.~C., {Allen}, G.~E., {et~al.} 2006, in Society
  of Photo-Optical Instrumentation Engineers (SPIE) Conference Series, Vol.
  6270, \procspie, 62701V

\bibitem[{{Gabriel} {et~al.}(2004){Gabriel}, {Denby}, {Fyfe}, {Hoar}, {Ibarra},
  {Ojero}, {Osborne}, {Saxton}, {Lammers}, \& {Vacanti}}]{sas04}
{Gabriel}, C., {Denby}, M., {Fyfe}, D.~J., {et~al.} 2004, in Astronomical
  Society of the Pacific Conference Series, Vol. 314, Astronomical Data
  Analysis Software and Systems (ADASS) XIII, ed. F.~{Ochsenbein}, M.~G.
  {Allen}, \& D.~{Egret}, 759

\bibitem[{{Gaia Collaboration}(2018)}]{gaia18}
{Gaia Collaboration}. 2018, VizieR Online Data Catalog, 1345

\bibitem[{{Harrison} {et~al.}(2013){Harrison}, {Craig}, {Christensen},
  {Hailey}, {Zhang}, {Boggs}, {Stern}, {Cook}, {Forster}, {Giommi},
  {Grefenstette}, {Kim}, {Kitaguchi}, {Koglin}, {Madsen}, {Mao}, {Miyasaka},
  {Mori}, {Perri}, {Pivovaroff}, {Puccetti}, {Rana}, {Westergaard}, {Willis},
  {Zoglauer}, {An}, {Bachetti}, {Barri{\`e}re}, {Bellm}, {Bhalerao},
  {Brejnholt}, {Fuerst}, {Liebe}, {Markwardt}, {Nynka}, {Vogel}, {Walton},
  {Wik}, {Alexander}, {Cominsky}, {Hornschemeier}, {Hornstrup}, {Kaspi},
  {Madejski}, {Matt}, {Molendi}, {Smith}, {Tomsick}, {Ajello}, {Ballantyne},
  {Balokovi{\'c}}, {Barret}, {Bauer}, {Blandford}, {Brandt}, {Brenneman},
  {Chiang}, {Chakrabarty}, {Chenevez}, {Comastri}, {Dufour}, {Elvis}, {Fabian},
  {Farrah}, {Fryer}, {Gotthelf}, {Grindlay}, {Helfand}, {Krivonos}, {Meier},
  {Miller}, {Natalucci}, {Ogle}, {Ofek}, {Ptak}, {Reynolds}, {Rigby},
  {Tagliaferri}, {Thorsett}, {Treister}, \& {Urry}}]{har13}
{Harrison}, F.~A., {Craig}, W.~W., {Christensen}, F.~E., {et~al.} 2013, \apj,
  770, 103, \dodoi{10.1088/0004-637X/770/2/103}

\bibitem[{(Heasarc)(2014)}]{heasarc14}
(Heasarc), N. H. E. A. S. A. R.~C. 2014, {HEAsoft: Unified Release of FTOOLS
  and XANADU}.
\newblock \doeprint{1408.004}

\bibitem[{{Ho} {et~al.}(2017){Ho}, {Ng}, {Lyne}, {Stappers}, {Coe}, {Halpern},
  {Johnson}, \& {Steele}}]{ho17}
{Ho}, W.~C.~G., {Ng}, C.-Y., {Lyne}, A.~G., {et~al.} 2017, \mnras, 464, 1211,
  \dodoi{10.1093/mnras/stw2420}

\bibitem[{{Johnson} {et~al.}(2018){Johnson}, {Wood}, {Kerr}, {Corbet},
  {Cheung}, {Ray}, \& {Omodei}}]{j18}
{Johnson}, T.~J., {Wood}, K.~S., {Kerr}, M., {et~al.} 2018, \apj, 863, 27,
  \dodoi{10.3847/1538-4357/aad185}

\bibitem[{{Kalberla} {et~al.}(2005){Kalberla}, {Burton}, {Hartmann}, {Arnal},
  {Bajaja}, {Morras}, \& {P{\"o}ppel}}]{kal05}
{Kalberla}, P.~M.~W., {Burton}, W.~B., {Hartmann}, D., {et~al.} 2005, \aap,
  440, 775, \dodoi{10.1051/0004-6361:20041864}

\bibitem[{{Khangulyan} {et~al.}(2011){Khangulyan}, {Aharonian}, {Bogovalov}, \&
  {Rib{\'o}}}]{khan11}
{Khangulyan}, D., {Aharonian}, F.~A., {Bogovalov}, S.~V., \& {Rib{\'o}}, M.
  2011, \apj, 742, 98, \dodoi{10.1088/0004-637X/742/2/98}

\bibitem[{{Khangulyan} {et~al.}(2012){Khangulyan}, {Aharonian}, {Bogovalov}, \&
  {Rib{\'o}}}]{khan12}
---. 2012, \apjl, 752, L17, \dodoi{10.1088/2041-8205/752/1/L17}

\bibitem[{{Kirk} {et~al.}(1999){Kirk}, {Ball}, \& {Skj{\ae}raasen}}]{kirk99}
{Kirk}, J.~G., {Ball}, L., \& {Skj{\ae}raasen}, O. 1999, Astroparticle Physics,
  10, 31, \dodoi{10.1016/S0927-6505(98)00041-3}

\bibitem[{{Kolka} {et~al.}(2017){Kolka}, {Eenm{\"a}e}, {Laur}, \&
  {Aret}}]{kolka17}
{Kolka}, I., {Eenm{\"a}e}, T., {Laur}, J., \& {Aret}, A. 2017, Research Notes
  of the American Astronomical Society, 1, 37, \dodoi{10.3847/2515-5172/aa9f17}

\bibitem[{{Li} {et~al.}(2014){Li}, {Torres}, \& {Zhang}}]{li14}
{Li}, J., {Torres}, D.~F., \& {Zhang}, S. 2014, \apjl, 785, L19,
  \dodoi{10.1088/2041-8205/785/1/L19}

\bibitem[{{Li} {et~al.}(2017{\natexlab{a}}){Li}, {Kong}, {Tam}, {Hou},
  {Takata}, \& {Hui}}]{li17a}
{Li}, K.~L., {Kong}, A.~K.~H., {Tam}, P.~H.~T., {et~al.} 2017{\natexlab{a}},
  \apj, 843, 85, \dodoi{10.3847/1538-4357/aa784e}

\bibitem[{{Li} {et~al.}(2018){Li}, {Takata}, {Ng}, {Kong}, {Tam}, {Hui}, \&
  {Cheng}}]{li18}
{Li}, K.~L., {Takata}, J., {Ng}, C.~W., {et~al.} 2018, \apj, 857, 123,
  \dodoi{10.3847/1538-4357/aab848}

\bibitem[{{Li} {et~al.}(2017{\natexlab{b}}){Li}, {Kong}, {Takata}, {Tam},
  {Cheng}, {He}, {Hui}, {Ng}, \& {Pal}}]{li17c}
{Li}, K.~L., {Kong}, A.~K.~H., {Takata}, J., {et~al.} 2017{\natexlab{b}}, The
  Astronomer's Telegram, 10993

\bibitem[{{Lyne} {et~al.}(2015){Lyne}, {Stappers}, {Keith}, {Ray}, {Kerr},
  {Camilo}, \& {Johnson}}]{lyne15}
{Lyne}, A.~G., {Stappers}, B.~W., {Keith}, M.~J., {et~al.} 2015, \mnras, 451,
  581, \dodoi{10.1093/mnras/stv236}

\bibitem[{{Petropoulou} {et~al.}(2018){Petropoulou}, {Vasilopoulos},
  {Christie}, {Giannios}, \& {Coe}}]{petro18}
{Petropoulou}, M., {Vasilopoulos}, G., {Christie}, I.~M., {Giannios}, D., \&
  {Coe}, M.~J. 2018, \mnras, 474, L22, \dodoi{10.1093/mnrasl/slx185}

\bibitem[{{Str{\"u}der} {et~al.}(2001){Str{\"u}der}, {Briel}, {Dennerl},
  {Hartmann}, {Kendziorra}, {Meidinger}, {Pfeffermann}, {Reppin}, {Aschenbach},
  {Bornemann}, {Br{\"a}uninger}, {Burkert}, {Elender}, {Freyberg}, {Haberl},
  {Hartner}, {Heuschmann}, {Hippmann}, {Kastelic}, {Kemmer}, {Kettenring},
  {Kink}, {Krause}, {M{\"u}ller}, {Oppitz}, {Pietsch}, {Popp}, {Predehl},
  {Read}, {Stephan}, {St{\"o}tter}, {Tr{\"u}mper}, {Holl}, {Kemmer}, {Soltau},
  {St{\"o}tter}, {Weber}, {Weichert}, {von Zanthier}, {Carathanassis}, {Lutz},
  {Richter}, {Solc}, {B{\"o}ttcher}, {Kuster}, {Staubert}, {Abbey}, {Holland},
  {Turner}, {Balasini}, {Bignami}, {La Palombara}, {Villa}, {Buttler},
  {Gianini}, {Lain{\'e}}, {Lumb}, \& {Dhez}}]{str01}
{Str{\"u}der}, L., {Briel}, U., {Dennerl}, K., {et~al.} 2001, \aap, 365, L18,
  \dodoi{10.1051/0004-6361:20000066}

\bibitem[{{Takahashi} {et~al.}(2009){Takahashi}, {Kishishita}, {Uchiyama},
  {Tanaka}, {Yamaoka}, {Khangulyan}, {Aharonian}, {Bosch-Ramon}, \&
  {Hinton}}]{hasi09}
{Takahashi}, T., {Kishishita}, T., {Uchiyama}, Y., {et~al.} 2009, \apj, 697,
  592, \dodoi{10.1088/0004-637X/697/1/592}

\bibitem[{{Takata} {et~al.}(2017){Takata}, {Tam}, {Ng}, {Li}, {Kong}, {Hui}, \&
  {Cheng}}]{takata17}
{Takata}, J., {Tam}, P.~H.~T., {Ng}, C.~W., {et~al.} 2017, \apj, 836, 241,
  \dodoi{10.3847/1538-4357/aa5c80}

\bibitem[{{Takata} {et~al.}(2012){Takata}, {Okazaki}, {Nagataki}, {Naito},
  {Kawachi}, {Lee}, {Mori}, {Hayasaki}, {Yamaguchi}, \& {Owocki}}]{takata12}
{Takata}, J., {Okazaki}, A.~T., {Nagataki}, S., {et~al.} 2012, \apj, 750, 70,
  \dodoi{10.1088/0004-637X/750/1/70}

\bibitem[{{Tam} {et~al.}(2018){Tam}, {He}, {Pal}, \& {Cui}}]{tam18}
{Tam}, P.~H.~T., {He}, X.-B., {Pal}, P.~S., \& {Cui}, Y. 2018, \apj, 862, 165,
  \dodoi{10.3847/1538-4357/aacf00}

\bibitem[{{Tam} {et~al.}(2015){Tam}, {Li}, {Takata}, {Okazaki}, {Hui}, \&
  {Kong}}]{tam15}
{Tam}, P.~H.~T., {Li}, K.~L., {Takata}, J., {et~al.} 2015, \apjl, 798, L26,
  \dodoi{10.1088/2041-8205/798/1/L26}

\bibitem[{{Tavani} \& {Arons}(1997)}]{tavani97}
{Tavani}, M., \& {Arons}, J. 1997, \apj, 477, 439, \dodoi{10.1086/303676}

\bibitem[{{Turner} {et~al.}(2001){Turner}, {Abbey}, {Arnaud}, {Balasini},
  {Barbera}, {Belsole}, {Bennie}, {Bernard}, {Bignami}, {Boer}, {Briel},
  {Butler}, {Cara}, {Chabaud}, {Cole}, {Collura}, {Conte}, {Cros}, {Denby},
  {Dhez}, {Di Coco}, {Dowson}, {Ferrando}, {Ghizzardi}, {Gianotti}, {Goodall},
  {Gretton}, {Griffiths}, {Hainaut}, {Hochedez}, {Holland}, {Jourdain},
  {Kendziorra}, {Lagostina}, {Laine}, {La Palombara}, {Lortholary}, {Lumb},
  {Marty}, {Molendi}, {Pigot}, {Poindron}, {Pounds}, {Reeves}, {Reppin},
  {Rothenflug}, {Salvetat}, {Sauvageot}, {Schmitt}, {Sembay}, {Short},
  {Spragg}, {Stephen}, {Str{\"u}der}, {Tiengo}, {Trifoglio}, {Tr{\"u}mper},
  {Vercellone}, {Vigroux}, {Villa}, {Ward}, {Whitehead}, \& {Zonca}}]{tur01}
{Turner}, M.~J.~L., {Abbey}, A., {Arnaud}, M., {et~al.} 2001, \aap, 365, L27,
  \dodoi{10.1051/0004-6361:20000087}

\bibitem[{{Uchiyama} {et~al.}(2009){Uchiyama}, {Tanaka}, {Takahashi}, {Mori},
  \& {Nakazawa}}]{u09}
{Uchiyama}, Y., {Tanaka}, T., {Takahashi}, T., {Mori}, K., \& {Nakazawa}, K.
  2009, \apj, 698, 911, \dodoi{10.1088/0004-637X/698/1/911}

\bibitem[{{Weisskopf} {et~al.}(2000){Weisskopf}, {Tananbaum}, {Van Speybroeck},
  \& {O'Dell}}]{wei00}
{Weisskopf}, M.~C., {Tananbaum}, H.~D., {Van Speybroeck}, L.~P., \& {O'Dell},
  S.~L. 2000, in \procspie, Vol. 4012, X-Ray Optics, Instruments, and Missions
  III, ed. J.~E. {Truemper} \& B.~{Aschenbach}, 2--16

\bibitem[{{Wik} {et~al.}(2014){Wik}, {Hornstrup}, {Molendi}, {Madejski},
  {Harrison}, {Zoglauer}, {Grefenstette}, {Gastaldello}, {Madsen}, \&
  {Westergaard}}]{wik14}
{Wik}, D.~R., {Hornstrup}, A., {Molendi}, S., {et~al.} 2014, \apj, 792, 48,
  \dodoi{10.1088/0004-637X/792/1/48}

\bibitem[{{Wilms} {et~al.}(2000){Wilms}, {Allen}, \& {McCray}}]{wam00}
{Wilms}, J., {Allen}, A., \& {McCray}, R. 2000, \apj, 542, 914,
  \dodoi{10.1086/317016}

\end{thebibliography}
% Don't change these lines
%\bsp    % typesetting comment
\label{lastpage}

\end{document}